\documentclass[%
 aip,
 amsmath,amssymb,
 reprint,%
]{revtex4-1}

\usepackage{graphicx}
\usepackage{dcolumn}
\usepackage{bm}

\usepackage[utf8]{inputenc}
\usepackage[T1]{fontenc}
\usepackage[nottoc]{tocbibind}
\usepackage{mathptmx}
\usepackage{etoolbox}
\usepackage{gensymb}
\usepackage{booktabs}

\makeatletter
\def\@email#1#2{%
 \endgroup
 \patchcmd{\titleblock@produce}
  {\frontmatter@RRAPformat}
  {\frontmatter@RRAPformat{\produce@RRAP{*#1\href{mailto:#2}{#2}}}\frontmatter@RRAPformat}
  {}{}
}%
\makeatother
\begin{document}

\preprint{AIP/123-QED}

\title[]{Magnetostatics of Room Temperature Compensated Co/Gd/Co/Gd-based Synthetic Ferrimagnets}
\author{Thomas J. Kools}
 \email{t.j.kools@tue.nl}
 \author{Marnix C. van Gurp}
\affiliation{%
Department of Applied Physics, Eindhoven University of Technology \\
P. O. Box 513, 5600 MB Eindhoven, The Netherlands
}%
\author{Bert Koopmans}%
\affiliation{%
Department of Applied Physics, Eindhoven University of Technology \\
P. O. Box 513, 5600 MB Eindhoven, The Netherlands
}%
\author{Reinoud Lavrijsen}
\affiliation{%
Department of Applied Physics, Eindhoven University of Technology \\
P. O. Box 513, 5600 MB Eindhoven, The Netherlands
}%

\date{\today}

\begin{abstract}
Flexibility for interface engineering, and access to all-optical switching of the magnetization, make synthetic ferrimagnets an interesting candidate for advanced opto-spintronic devices. Moreover, due to their layered structure and disordered interfaces they also bear promise for the emerging field of graded magnetic materials. The fastest and most efficient spin-orbit torque driven manipulation of the magnetic order in this material system generally takes place at compensation. Here, we present a systematic experimental and modeling study of the conditions for magnetization compensation and perpendicular magnetic anisotropy in the synthetic ferrimagnetic Co/Gd/Co/Gd system. A model based on partial intermixing at the Co/Gd interfaces of this system has been developed which explains the experiments well, and provides a new tool to understand its magnetic characteristics. More specifically, this work provides new insight in the decay of the Co proximity-induced magnetization in the Gd, and the role the capping layer plays in the Gd magnetization. 
\end{abstract}

\maketitle

The ever expanding rate of data generation and consumption propels research into new material systems to use for processing and storage of information. Therefore, one major challenge of contemporary research in spintronics is to develop material systems of which the magnetization can be manipulated both time- and energy-efficiently. 3d-4f ferrimagnetic material systems, like GdFeCo and CoTb alloys, and multilayers based on a combination of these metals, are attractive due to their antiferromagnetically coupled sublattices \cite{Kim2022,Kim2017,Caretta2018,Sala2022,vanHees2020,Wang2022,Cai2020}. These materials aim to combine favorable properties of their ferromagnetic and antiferromagnetic counterparts, and bear promise for the emerging field of graded magnetism \cite{Fallarino_2021}. They have garnered a great amount of attention from the scientific community due to their access to single-pulse all-optical switching (AOS) of the magnetization \cite{Radu2011, Li:2021wr, Lalieu2017AOS, vanHees2020}, efficient spin-orbit torque (SOT)-driven manipulation of the magnetic order \cite{JeSOT,UedaSOT,FinleySOT,Mishra2017,Roschewsky2017} and exchange torque driven current induced-domain wall motion (CIDWM) with velocities over 1000 m/s \cite{Caretta2018,Cai2020,Kim2017}. Hence, these developments push the search for material platforms for domain wall-based memory in advanced solid state devices like racetrack memory \cite{parkin2008magnetic,Blasing2020review,Pham2016}. Interestingly, the combination of AOS and efficient CIDWM in this material system is also very promising to bridge the gap between photonics and spintronics \cite{Lalieu2019,Li:2022CIDWM,2022Luding,Demirer2022}.

Co/Gd-based synthetic ferrimagnetic bilayers, where the 3d and 4f-material are grown as discrete layers, have a few distinct advantages over 3d-4f alloys. The layered structure of these synthetic ferrimagnets allows for easier adaption to wafer scale production. Also, contrary to alloys, a much wider composition range between the 3d and 4f-metal exhibits AOS \cite{Beens2019Comparing,Beens2019Intermixing}. Combined with the increased access to interfacial engineering, this leads to more flexibility and tunability of its magnetic properties. Moreover, the Pt/Co/Gd trilayer displays strong interfacial spintronic effects, such as perpendicular magnetic anisotropy (PMA), the spin-Hall effect, and the interfacial Dzyaloshinskii–Moriya interaction, all important aspects for applications based on efficient domain wall motion \cite{Blasing2018, Pham2016,Ryu2013,Yang2015a}. Despite these favorable properties, the engineering relevance of the Co/Gd bilayer system has been limited due to the absence of both magnetization and angular momentum compensation, where the two magnetic sublattices cancel each other, at room temperature. For it is well known that CIDWM \cite{Caretta2018, Kim2017,Li:2022CIDWM} and in general SOT-driven ferrimagnetic spin dynamics \cite{JeSOT,UedaSOT,FinleySOT,Sala2022} are most effective close to the angular momentum or magnetization compensation point. 



\begin{figure}
\includegraphics[width=0.47\textwidth]{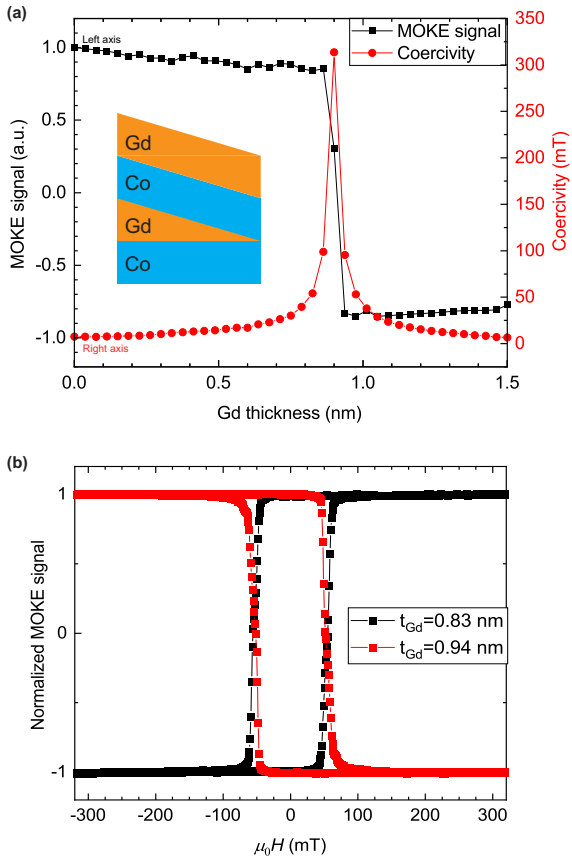}
\caption{\label{fig:wide}\label{fig:Figure1} Polar MOKE characterization of a Co/Gd/Co/Gd sample where the middle Gd layer is wedged between 0 and 1.5 nm. a): Remanent polar MOKE signal normalized by its value at $t_\mathrm{Gd1}$ = 0 nm (black), and coercive field (red) as a function of Gd layer thickness. b): Two sample hysteresis loops measured by polar MOKE at a Co (black) and Gd-dominated magnetic composition (red) on the wedge in a).}
\end{figure}

\begin{figure}
\includegraphics[width=0.47\textwidth]{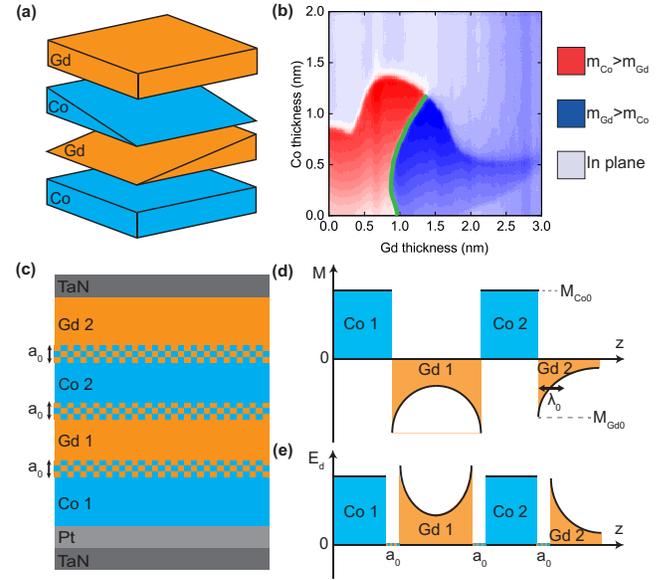}
\caption{\label{fig:Figure2} a): Schematic of the double wedge samples under investigation in this report. b): Polar MOKE scan of a double wedge Co(0.6 nm)/Gd(x-axis)/Co(y-axis)/Gd sample(1.5). The compensation boundary is indicated by the green line. c): Schematic illustration of the model used to describe the compensation and SRT-boundary for the magnetostatic phase diagrams. Magnetic layers are modelled with the inclusion of an intermixing region with a width $a_\mathrm{0}$. d) and e) respectively, illustrate the magnetization and shape anisotropy energy as modelled throughout the four magnetic layers.}
\end{figure}

In this work, we therefore investigate the conditions for compensation in Co/Gd/Co/Gd, which we from now on dub the quadlayer system. 
Compared to the Co/Gd bilayer, we double the magnetic volume of the Co while tripling the number of Co/Gd interfaces where magnetization is induced in the Gd through direct exchange with the Co\cite{Lalieu2017AOS,Pham2016}. 
This is expected to enhance the contribution of the Gd to the net magnetic moment, while still maintaining PMA. 
The samples nominally consist of stacks of TaN(4 nm)/Pt(4)/Co(0.6)/Gd($t_\mathrm{Gd1}$)/Co($t_\mathrm{Co2}$)/Gd($t_\mathrm{Gd2}$)/TaN(4) as schematically drawn in Fig.\ref{fig:Figure2}c, which were grown on Si/SiO$_\mathrm{2}$ substrates through magnetron sputtering in a chamber with a typical base pressure of $5 \times 10^{-9}$ mBar. 
The first sample is fabricated using wedge sputtering in order to confirm that compensation is achieved. Specifically, in the first sample the middle Gd thickness $t_\mathrm{Gd1}$ is varied between 0 and 1.5 nm over a few mm (see inset Fig. \ref{fig:Figure1}a), whereas $t_\mathrm{Co2}$ and $t_\mathrm{Gd2}$ are constant and set to 0.7 and 1.5 nm, respectively. 

The magnetic properties of this wedge were investigated by the polar magneto-optic Kerr effect (pMOKE), where we are only sensitive to out-of-plane (OOP) components of the Co magnetization, as Gd does not contribute appreciably to the pMOKE signal at our used wavelength of 658 nm \cite{PhysRevB.8.1239}. We scan the sample locally using a focused laser spot. At magnetic compensation (e.g. from a Co-dominated to a Gd dominated region or vice-versa) two effects are expected: a divergence of the coercivity and a sign change in the pMOKE signal. The former can be observed in Fig. \ref{fig:Figure1}a, where the coercivity extracted from hysteresis loops measured across the wedge is plotted in red. The divergence follows from the inefficiency of the Zeeman interaction in a compensated system. This divergence coincides with a change in sign of the remanent pMOKE signal (Kerr rotation, normalized to its value at $t_\mathrm{Gd1}$= 0 nm) which is plotted in black in Fig. \ref{fig:Figure1}a. To understand this sign change, we must consider that in the Gd-dominated regime the Zeeman energy dictates that the Gd magnetization aligns with the magnetic field. The measured Co-magnetization will consequently align antiparallel to the field, leading to the change in sign of the pMOKE signal. The change in the hysteresis is illustrated in Fig. \ref{fig:Figure1}b, where the black and red loops are measured in the Co ($t_\mathrm{Gd1}$= 0.83 nm) and Gd-dominated magnetic regime ($t_\mathrm{Gd1}$= 0.94 nm), respectively. The 100\% remanence observed indicates the PMA in this sample. 


In order to obtain information on the tunability of the compensation point and PMA, as a low net magnetization would imply a large effective anisotropy, we use orthogonal double wedge samples. In Fig. \ref{fig:Figure2}a we illustrate the Co/Gd/Co/Gd double wedge sample structure. After deposition of the first Gd wedge, the sample is rotated by 90 degrees and the Co wedge is deposited. After the sample is saturated with an OOP magnetic field of 1 T we scan the sample surface and determine the remanence from the pMOKE signal at each point when no magnetic field is applied. Using this method allows us to scan the full parameter space of nominal layer thicknesses of the middle two layers in a single sample. A typical resulting diagram of the remanent pMOKE signal is shown in Fig. \ref{fig:Figure2}b for a sample where $t_\mathrm{Gd1}=0-3$ nm and $t_\mathrm{Co2}=0-2$ nm, keeping the top Gd thickness at $t_\mathrm{Gd2}$ = 1.5 nm at which we anticipate, based on earlier work, the proximity-induced magnetization in the Gd to be saturated  \cite{Pham2016,Lalieu2017AOS}. In the diagram, we can distinguish between three basic states. The red and dark blue regions indicate stack compositions where the magnetization points OOP, with the Co or Gd magnetization being dominant, respectively. The light blue region indicates stack compositions where the magnetization points in-plane (IP); this is above the spin reorientation transition (SRT), where the interfacial PMA is not sufficient to keep the full stack OOP. These three regions define two major transitions of interest: the compensation boundary (red to dark blue) and the SRT boundary (red/dark blue to light blue).


To obtain a quantitative understanding of the shapes of these boundaries, a model has been developed to simultaneously describe the compensation boundary and the SRT boundary. Furthermore, the model will be used to get insight in the basic properties of the proximity induced magnetization in the Gd and level of intermixing, a quantity that has not been investigated using these double wedged samples. We set out to model the net magnetization, which is zero at the compensation boundary, as well as the magnetostatic free energy of the anisotropy, which is zero at the SRT boundary. One of the main assumptions of the model relies on the experimental observation that the interface between Co and Gd thin films are intermixed \cite{Andres, Alonso_2002, Nishimura2020TEM}. The assumed Co and Gd concentration as a function of thickness is illustrated in Fig. \ref{fig:Figure2}c. The layers are modelled by means of the four magnetic layers in our Co/Gd/Co/Gd structure, with each layer assumed to be separated by an intermixing region with a constant and identical width of $a_\mathrm{0}$.

In order to find the magnetization compensation point, we then describe the net magnetization of this multilayer structure, which vanishes at compensation. We use typical assumptions for the magnetization profile of the Co/Gd bilayer to describe the magnetization in our Co/Gd/Co/Gd system, which are illustrated in Fig. \ref{fig:Figure2}d \cite{Pham2016,Lalieu2017AOS}. The Co magnetization is crudely assumed constant throughout the nominal thickness of the Co layer, with a value $M_\mathrm{Co0}$, giving: 

\begin{equation}
    M_\mathrm{Co1}=M_\mathrm{Co0},
\end{equation}
\noindent and
\begin{equation}
    M_\mathrm{Co2}=M_\mathrm{Co0}F_\mathrm{eq,Co},
\end{equation}

\noindent where, in order to implement the intermixing regions into the change of the magnetization with layer thickness, we empirically define the continuous function $F_\mathrm{eq,Co}$  (see Sup. \ref{Appendix: Perco} for details). It describes the transition to an equilibrium magnetic state with middle Co layer thickness caused both by the effect of intermixing on the magnetization, as well as percolative behavior. The latter of which describes the minimum thickness needed to stabilize a coherent ferromagnetic state.

In contrast to the Co magnetization, the magnetization in the Gd layers is mainly induced at the interface with the Co layer  \cite{Lalieu2017AOS, Pham2016}. Therefore, the magnetization as a function of the distance to the Co/Gd interface $z^*$ will be described by an exponentially decaying profile, which is typical to describe magnetization induced at an interface between a ferromagnet and a non-magnetic metal \cite{Swindells2022PIM,DIDRICHSEN1999Exponential,Demirer2021}, given by:

\begin{equation}
    M_\mathrm{Gd1}=M_\mathrm{Gd0}(\exp \left( -\frac{z^*}{\lambda_\mathrm{0}} \right) + \exp \left( \frac{z^*-t_\mathrm{Gd1}}{\lambda_\mathrm{0}} \right))F_\mathrm{eq,Co}F_\mathrm{eq,Gd},
\end{equation}
and
\begin{equation}
    M_\mathrm{Gd2}= M_\mathrm{Gd0}\exp \left( -z^*/\lambda_\mathrm{0} \right)F_\mathrm{eq,Co}F_\mathrm{eq,Gd},
\end{equation}

\noindent where $M_\mathrm{Gd0}$ is the magnitude of the magnetization at the interface, $\lambda_\mathrm{0}$ is the characteristic decay length of the magnetization, and $F_\mathrm{eq,Gd}$ is a similar empirical function to $F_\mathrm{eq,Co}$, describing the development of the Gd magnetization with middle Gd layer thickness (see Sup. \ref{Appendix: Perco}). Note that the effective exponential decay constant $\lambda_\mathrm{0}$ is influenced by many parameters, like surface roughness, the actual degree of intermixing, local ratios between Co and Gd atoms and the actual decay of the magnetization induced in the Gd, and should hence be interpreted as an effective parameter describing the collective behavior of all these effects. The resulting total magnetic moment per unit area $m_\mathrm{tot}$ can then be extracted by integrating the magnetizations over the respective layer thicknesses:

\begin{equation}
   m_\mathrm{tot}=\sum_{i=1}^{4}  \int_{0}^{t_\mathrm{i}} M_\mathrm{i} \,dz^*.
\end{equation}


\begin{figure*}
\centering
\includegraphics[width=1.07\textwidth]{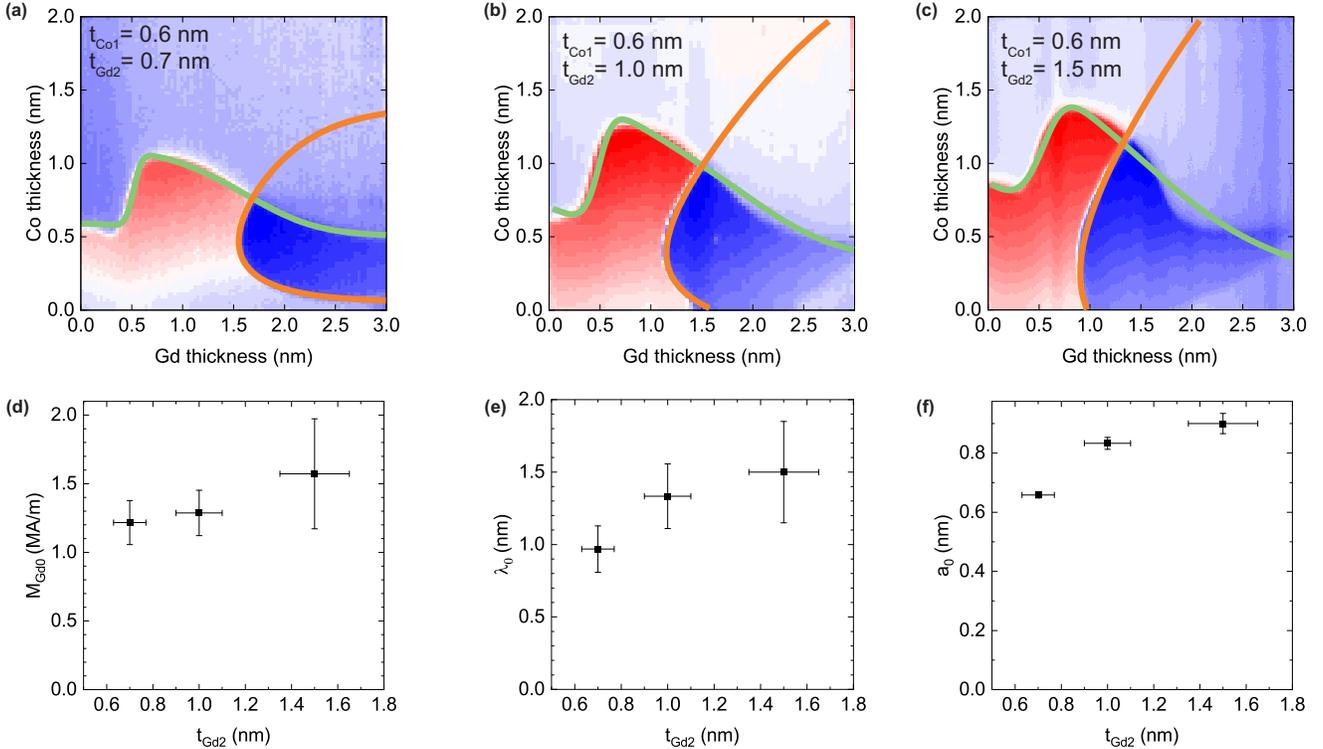}
\caption{\label{fig:Figure4} Magnetostatic phase diagrams with model fits of the magnetization (orange) and demagnetization energy (green) for $t_\mathrm{Gd2} =$ a): 0.7 nm, b): 1.0 nm, and c): 1.5 nm. The main magnetic parameters extracted from the fitting procedure for the three phase diagrams: d): Magnetization of Gd at the Co/Gd interface $M_\mathrm{Gd0}$, e): Magnetization decay length $\lambda_\mathrm{0}$, and f): intermixing region width $a_\mathrm{0}$.}
\end{figure*}


Next, the SRT-condition needs to be implemented. There are two main contributions to the effective anisotropy: the interfacial anisotropy energy and demagnetization energy. The magnetocrystalline anisotropy energy per unit area $K_\mathrm{S}$ due to the Co/Pt interface is assumed constant. The free energy density per unit area due to the shape anisotropy $E_\mathrm{d}$ is schematically plotted in Fig. \ref{fig:Figure2}e. It is calculated by treating the system as a continuous magnetic system and integrating the typical expression for the volume demagnetization energy density of a thin film with OOP magnetization $M_\mathrm{S}$: $E_\mathrm{d}^* = \frac{1}{2} \mu_\mathrm{0} M_\mathrm{S}^2$, where $\mu_\mathrm{0}$ is the magnetic permeability of vacuum. In order to account for the smaller demagnetizing field in the intermixing regions where the magnetization is inherently lower, we subtract the demagnetization energy of the intermixing region, with a characteristic width of $a_\mathrm{0}$, from the total demagnetization energy leading to the following expression for the total area-normalized demagnetization energy $E_\mathrm{d}$:

\begin{eqnarray}
 E_\mathrm{d}=&& \sum_{i=1}^{4}  \int_{0}^{t_\mathrm{i}} \frac{1}{2} \mu_\mathrm{0} M_\mathrm{i}^2 \,dz^*\nonumber\\
  && -\biggl( \int_{0}^{a_\mathrm{0}/2} \frac{1}{2} \mu_\mathrm{0} M_\mathrm{Co1}^2 \,dz^* + F_\mathrm{eq,mix} \int_{0}^{a_\mathrm{0}} \frac{1}{2} \mu_\mathrm{0} M_\mathrm{Gd1}^2 \,dz^* \nonumber \\
  &&+ F_\mathrm{eq,mix} \int_{0}^{a_\mathrm{0}} \frac{1}{2} \mu_\mathrm{0} M_\mathrm{Co2}^2 \,dz^* \nonumber \\
  &&+ \int_{0}^{a_\mathrm{0}/2} \frac{1}{2} \mu_\mathrm{0} M_\mathrm{Gd2}^2 \,dz^* \biggr),
\end{eqnarray}

where $F_\mathrm{eq,mix}$ is an identical empirical function to 
$F_\mathrm{eq,Gd}$ and $F_\mathrm{eq,Co}$ used to describe the onset and saturation of the intermixing regions in the Gd layer upon changing the layer thickness (see Sup. \ref{Appendix: Perco}). The resulting total free energy density per unit area is $E_\mathrm{tot}$ can then be calculated by adding $E_\mathrm{d}$ and $K_\mathrm{S}$ together.

We will use this model for the magnetization and anisotropy energy to test our physical understanding and make an estimate of the (effective) physical parameters underpinning these systems by fitting it to the measured phase diagrams, again considering the SRT-boundary and compensation boundary to be at $E_\mathrm{tot}$ = 0 and $m_\mathrm{tot}$ = 0, respectively.
To test the quantitative applicability of our model to these Co/Gd/Co/Gd systems, three double-wedged samples are considered: Co(0.6)/Gd(0-3)/Co(0-2)/Gd($t_\mathrm{Gd2}$) with $t_\mathrm{Gd2}$= 0.7, 1.0 and 1.5 nm, respectively, where the thicknesses chosen are expected to probe different degrees of decay of the induced magnetization in the Gd. The phase diagrams measured on the three samples are shown in Fig. \ref{fig:Figure4} a, b and c for  $t_\mathrm{Gd2}$ = 0.7, 1.0 and 1.5 nm, respectively, where we can immediately observe that with increasing top Gd thickness the area of the Gd-dominated region increases. 

Before fitting the model to explain the features of these phase diagrams, we experimentally characterize the interfacial anisotropy strength due to the Co/Pt interface $K_\mathrm{S}$ to be 1.22 mJ/m$^2$ (see Appendix \ref{appendix: anisotropy}), and set the Co saturation magnetization $M_\mathrm{Co0}$ equal to the bulk magnetization of Co at 1.4 MA/m. The other parameters in the model are left unconstrained. In Fig. \ref{fig:Figure4}a,b,c we give the resulting fits for $m_\mathrm{tot} = 0$ and $E_\mathrm{d} = 0$ in orange and green, respectively. We find a good correspondence between the model and the experiment. Specifically, The curvature of the magnetization profile, and the corresponding magnetostatic energy balance characterized by the peaked shape, are both generally well described. Particularly for the samples with $t_\mathrm{Gd2}$= 0.7 and 1 nm the correspondence is good across the whole phase diagrams. For $t_\mathrm{Gd2}$= 1.5 nm in Fig. \ref{fig:Figure4}c, the model correspondence on the SRT-boundary becomes worse for $t_\mathrm{Gd1}>$1.5 nm. It is not yet been unequivocally established what causes this difference between experiment and theory. We speculate that it might be due to finite size effects affecting the Curie temperature in the Gd and hence the total amount of magnetization induced beyond what is currently implemented in the model. Based on the model we can attribute the peak in effective anisotropy in the phase diagrams to the first $\sim$ 1 nm of Gd contributing mainly to the intermixing regions, leading to the initial increase, after which pure Gd is found, which decreases the effective anisotropy, leading to the decline from $\sim$ 1 nm onwards. Moreover, the change in curvature of the compensation boundary for $t_\mathrm{Co}<0.5$ nm between the three samples can now be attributed to the percolation limit approach of the Co layer leading to either one or three interfaces which induce a net magnetization in the Gd layers. 

To also illustrate the quantitative value of the model, we will now discuss the magnetic parameters extracted from these fits. In particular, the parameters fixing the hitherto unknown Gd magnetization profile $\lambda_\mathrm{0}$, $M_\mathrm{Gd0}$, and $a_0$ are of interest here. These parameters are plotted for the three fitting procedures in Fig. \ref{fig:Figure4}d, e and f, respectively. All other parameters found in the fitting procedure are listed in appendix \ref{app:fittingparameters}. The extracted Gd interfacial magnetization of about 1.3 MA/m is comparable with values found in earlier work \cite{Blasing2018,Pham2016,Lalieu2017AOS}. Next, the $\lambda_\mathrm{0}$ values in the order of 1 nm suggest that the magnetization profile extends well beyond the first monolayer affected by direct exchange with the Co layer. This observation is further corroborated when considering the difference between identical phase diagrams capped with Ta and TaN (see Appendix \ref{appendix: cap}). There we observe that the Gd- dominated OOP regime (dark blue in earlier diagrams) in the phase diagram extends all the way to Co-thicknesses of 0 nm in the TaN-capped sample, whereas in Ta-capped samples of otherwise identical composition a minimum middle Co thickness is always required to reach the Gd-dominated regime, indicating an overall reduction in the total magnetization of the Gd. We postulate that this difference is caused by magnetization quenching in the Gd due to intermixing between the capping layer and the Gd, a process that will likely be more severe for atomic Ta than for covalently bound TaN \cite{OKU1996265}. Finally, we will discuss the resulting values for $a_\mathrm{0}$ (Fig. \ref{fig:Figure4}f). Regarding the growth of Gd on Co and vice versa, earlier work demonstrated that the interfaces between multilayers of Co and Gd are disordered \cite{Andres,Alonso_2002,QUIROS2012933}, and that the exact growth and intermixing dynamics also depend on the order of growth of the two layers \cite{Andres,clemens_bain_payne_hufnagel_brennan_1991}. This indicates that the found intermixing region width $a_\mathrm{0}$ of around 0.8 nm is in line with earlier work. A reasonable comparison can be made to the [Pt/Co/Gd]-multilayers investigated by Nishimura et al., where using transmission electron microscopy investigations similar typical intermixing region widths were found as we find from fitting our model here, i.e., in the 0.5-1 nm range \cite{Nishimura2020TEM}.

In conclusion, we have experimentally demonstrated magnetic compensation in the synthetic ferrimagnetic quadlayer Co/Gd/Co/Gd system. It is found that compensation can be effectively tuned by layer thickness. We also demonstrated the utility of orthogonally wedged samples to characterize the nominal thickness parameter space in order to investigate the magnetostatics of these systems and consequently find stack compositions with favorable magnetic properties. Finally, a crude model for the net magnetization and PMA was developed which described the experiments well, providing an effective framework to discuss the magnetostatics in these compensated multilayered ferrimagnetic systems with PMA. We note that this is probably an oversimplified model to describe the real intermixing profiles, e.g. the constant Co magnetization with thickness. It however provides a good qualitative framework to build more detailed models which will require refinement of the assumed magnetization profiles using high-resolution depth sensitive magnetometry.  This work improves the understanding of basic magnetostatic properties and gives insight in the more fundamental aspects of the design and physics of these promising and flexible multilayer systems.

\begin{acknowledgments}

This work was part of the research program Foundation for Fundamental Research on Matter (FOM) and Gravitation program “Research Center for Integrated Nanophotonics,” which are financed by the Dutch Research Council (NWO). This work was suported by the Hedrik Casimir Institute.

\end{acknowledgments}

\section*{Author Delcarations}

\subsection*{Conflict of Interest}
    The authors have no conflicts to disclose

\subsection*{Author Contriubtions}
\noindent
\textbf{Thomas J. Kools:} Conceptualization (equal); Investigation (lead);
Methodology (lead); Writing – original draft (lead); Writing – review
and editing (lead). \textbf{Bert Koopmans:} Conceptualization (equal);
Funding acquisition (equal); Supervision (equal); Writing – review
and editing (supporting). \textbf{Reinoud Lavrijsen:} Conceptualization
(equal); Funding acquisition (equal); Supervision (equal); Writing –
review and editing (supporting).
    
\section*{Data Availability}
The data that support the findings of this study are available from the corresponding author upon reasonable request.

\appendix

\newpage 



\clearpage

\section{Percolation functions}\label{Appendix: Perco}
In order to implement the intermixing regions into the change of the magnetization with layer thickness, we empirically define continuous functions $F_\mathrm{eq,Gd}$ and $F_\mathrm{eq,Co}$ describing the transition to an equilibrium state with respective layer thickness. These describe both percolative behavior; the minimum thickness needed to stabilize a coherent ferromagnetic state, and the effect of intermixing on particularly the Gd magnetization:
\begin{subequations}
\label{eq:Percolation}
\begin{equation}
    F_\mathrm{eq,Gd}(t_\mathrm{Gd1})=\frac{\mathrm{erf}\left( (t_\mathrm{Gd1}-t_\mathrm{0,Gd})/L_\mathrm{Gd1} \right)+1}{2},
\end{equation}
\begin{equation}
     F_\mathrm{eq,Co}(t_\mathrm{Co2})=\frac{\mathrm{erf}\left( (t_\mathrm{Co2}-t_\mathrm{0,Co})/L_\mathrm{Co2} \right)+1}{2},
\end{equation}
\end{subequations}
where $t_\mathrm{0,Gd1}$ and $t_\mathrm{0,Co2}$, and $L_\mathrm{Gd1}$ and $L_\mathrm{Co2}$, are parameters defining the critical thickness and characteristic width of the percolation, respectively.

$F_\mathrm{eq,mix}$ is an identical empirical function to those presented in Eq. \ref{eq:Percolation} used to describe the onset of the intermixing regions in the Gd layer upon changing the layer thickness: 
\begin{equation}
    F_\mathrm{eq,mix}(t_\mathrm{Gd1})=\frac{\mathrm{erf}\left( (t_\mathrm{Gd1}-t_\mathrm{0,mix})/L_\mathrm{mix} \right)+1}{2}.
\end{equation}

\section{Characterization Ks}\label{appendix: anisotropy}
\begin{figure}
\centering
\includegraphics[width=0.5\textwidth]{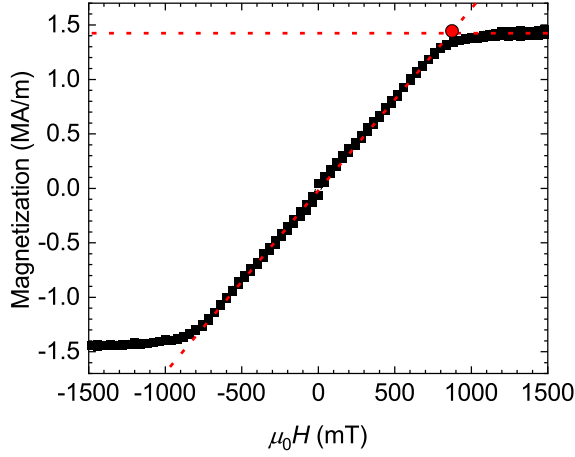}
\caption{\label{fig:anisotropy} In-plane SQUID characterization of the magnetic moment of a Ta(4 nm)/Pt(4)/Co(1)/TaN(4) as a function of in-plane field. The red dot indicates the extracted anisotropy field.}
\end{figure}

In order to estimate the anisotropy constant $K_\mathrm{S}$, we performed IP VSM-SQUID measurements (see Fig. \ref{fig:anisotropy}) on a Ta(4)/Pt(4)/Co(1)/TaN(4) sample. The resulting hard-axis response to the IP field is typical of a sample with PMA like this multilayer. We find an anisotropy field of 900 mT. To estimate the corresponding $K_\mathrm{S}$, we use a simple Stoner-Wolfarth theory, considering three contributions to the magnetostatic free energy: Zeeman energy $E_\mathrm{z}$, interfacial anisotropy from the Co/Pt interface $E_\mathrm{K}$, and the shape anisotropy $E_\mathrm{s}$. The resulting total energy $E_\mathrm{tot}$ is then given by the sum of these three contributions:
\begin{equation}
    E_\mathrm{tot}=\frac{1}{2}\mu_\mathrm{0}M_\mathrm{S}^2\mathrm{cos}^2\left(\theta\right)+\mathrm{sin}\left(\theta\right)\left(-\mu_\mathrm{0} H M_\mathrm{S}+\frac{K_\mathrm{S} \mathrm{sin}\left(\theta\right)}{t}\right),
\end{equation}
where $\mu_\mathrm{0}$ is the permeability of vacuum, $M_\mathrm{S}$ is the saturation magnetization, $\theta$ is the angle between the magnetization and the thin film sample normal, $H$ is the applied magnetic field, and $t$ is the thickness of the magnetic layer. By minimizing $E_\mathrm{tot}$ with respect to $\theta$ and setting $\theta=90\degree$ we find the anisotropy field $H_\mathrm{a}$:
\begin{equation}
    H_\mathrm{a}=\frac{1}{2}\mu_\mathrm{0}t\left(H M_\mathrm{S} + M_\mathrm{S}^2\right).
\end{equation}
For $M_\mathrm{S}=$1.4 MA/m (SQUID), $t=$1 nm and $H_\mathrm{a} =$ 900 mT (SQUID), we find $K_\mathrm{S}= 1.22 $ mJ/m$^2$.

\section{Fitting parameters}\label{app:fittingparameters}
The parameters for the best fit of the model described in the main text to the double wedge Co/Gd/Co/Gd samples as shown in Figs. \ref{fig:Figure4}a, \ref{fig:Figure4}b and \ref{fig:Figure4}c are described in this section. Tables \ref{tablefit0.7}, \ref{tablefit1.0} and \ref{tablefit1.5} show these fitting parameters for top Gd thickness $t_\mathrm{Gd2}=0.7$ nm, $t_\mathrm{Gd2}=1.0$ nm, $t_\mathrm{Gd2}=1.5$ nm, respectively.

\begin{table}[h]
\begin{center}
\begin{minipage}{0.47\textwidth}
\caption{Summary of the fitting parameters found to best describe the compensation boundary and boundary from OOP to IP magnetization in the double wedge with top Gd thickness $t_\mathrm{Gd2}=0.7$ nm (see Fig. \ref{fig:Figure4}a). For $M_\mathrm{S,Co}$ and $K_\mathrm{S}$ we choose 1.4 MA/m and 1.22 mJ/m$^2$ as explained in the main text and appendix \ref{appendix: anisotropy}. Errors represent 95\% confidence intervals extracted from the fitting procedure.}\label{tablefit0.7}
\begin{tabular}{@{}llll@{}}
\toprule
Parameter & Value & Error & Unit\\
\midrule
$M_\mathrm{Gd1}$     & 1.2 & 0.2 & MA/m  \\
$t_\mathrm{0,Co}$    & 0.21 & 0.01 & nm   \\
$t_\mathrm{0,Gd1}$    & 1.2 & 0.2 & nm    \\
$t_\mathrm{0,mix}$     & 0.47 & 0.01 & nm  \\
$L_\mathrm{Co}$    & 0.43 & 0.03 & nm    \\
$L_\mathrm{Gd1}$    & 0.83 & 0.02 & nm    \\
$L_\mathrm{mix}$     & 1.3 & 0.1 & nm   \\
$\lambda_\mathrm{0}$    & 0.97 & 0.16 & nm   \\
$a_\mathrm{0}$    & 0.66 & 0.01 & nm   \\
\botrule
\end{tabular}
\end{minipage}
\end{center}
\end{table}

\begin{table}[h]
\begin{center}
\begin{minipage}{0.47\textwidth}
\caption{Summary of the fitting parameters found to best describe the compensation boundary and boundary from OOP to IP magnetization in the double wedge with top Gd thickness $t_\mathrm{Gd2}=1.0$ nm (see Fig. \ref{fig:Figure4}b). For $M_\mathrm{S,Co}$ and $K_\mathrm{S}$ we choose 1.4 MA/m and 1.22 mJ/m$^2$ as explained in the main text and appendix \ref{appendix: anisotropy}. Errors represent 95\% confidence intervals extracted from the fitting procedure.}\label{tablefit1.0}
\begin{tabular}{@{}llll@{}}
\toprule
Parameter & Value & Error &  Unit\\
\midrule
$M_\mathrm{Gd1}$     & 1.29 & 0.17 & MA/m  \\
$t_\mathrm{0,Co}$    & 0.13  & 0.01 & nm   \\
$t_\mathrm{0,Gd1}$    & 0.72 & 0.15  & nm    \\
$t_\mathrm{0,mix}$     & 0.39 & 0.01 & nm  \\
$L_\mathrm{Co}$    & 0.63  & 0.02 & nm    \\
$L_\mathrm{Gd1}$    & 0.56 & 0.06 & nm     \\
$L_\mathrm{mix}$     & 0.27  & 0.01& nm  \\
$\lambda_\mathrm{0}$    & 1.3 & 0.2 & nm   \\
$a_\mathrm{0}$    & 0.83 & 0.02 & nm   \\

\botrule
\end{tabular}
\end{minipage}
\end{center}
\end{table}

\begin{table}[h]
\begin{center}
\begin{minipage}{0.47\textwidth}
\caption{Summary of the fitting parameters found to best describe the compensation boundary and boundary from OOP to IP magnetization in the double wedge with top Gd thickness $t_\mathrm{Gd2}=1.5$ nm (see Fig. \ref{fig:Figure4}c). For $M_\mathrm{S,Co}$ and $K_\mathrm{S}$ we choose 1.4 MA/m and 1.22 mJ/m$^2$ as explained in the main text and appendix \ref{appendix: anisotropy}. Errors represent 95\% confidence intervals extracted from the fitting procedure.}\label{tablefit1.5}
\begin{tabular}{@{}llll@{}}
\toprule
Parameter & Value &  Unit\\
\midrule
$M_\mathrm{Gd1}$     & 1.57 & 0.16 & MA/m  \\
$t_\mathrm{0,Co}$    & 0.10 & 0.02 & nm   \\
$t_\mathrm{0,Gd1}$    & 0.66 & 0.21  & nm    \\
$t_\mathrm{0,mix}$     & 0.43 & 0.01 & nm  \\
$L_\mathrm{Co}$    & 1.14 & 0.03 & nm    \\
$L_\mathrm{Gd1}$    & 3.1 & 1.1 & nm     \\
$L_\mathrm{mix}$     & 0.28 & 0.01 & nm   \\
$\lambda_\mathrm{0}$    & 1.50 & 0.35 & nm   \\
$a_\mathrm{0}$    & 0.90 & 0.03 & nm   \\
\botrule
\end{tabular}
\end{minipage}
\end{center}
\end{table}

\section{Comparison phase diagram capping layers}\label{appendix: cap}
\begin{figure}
\centering
\includegraphics[width=0.47\textwidth]{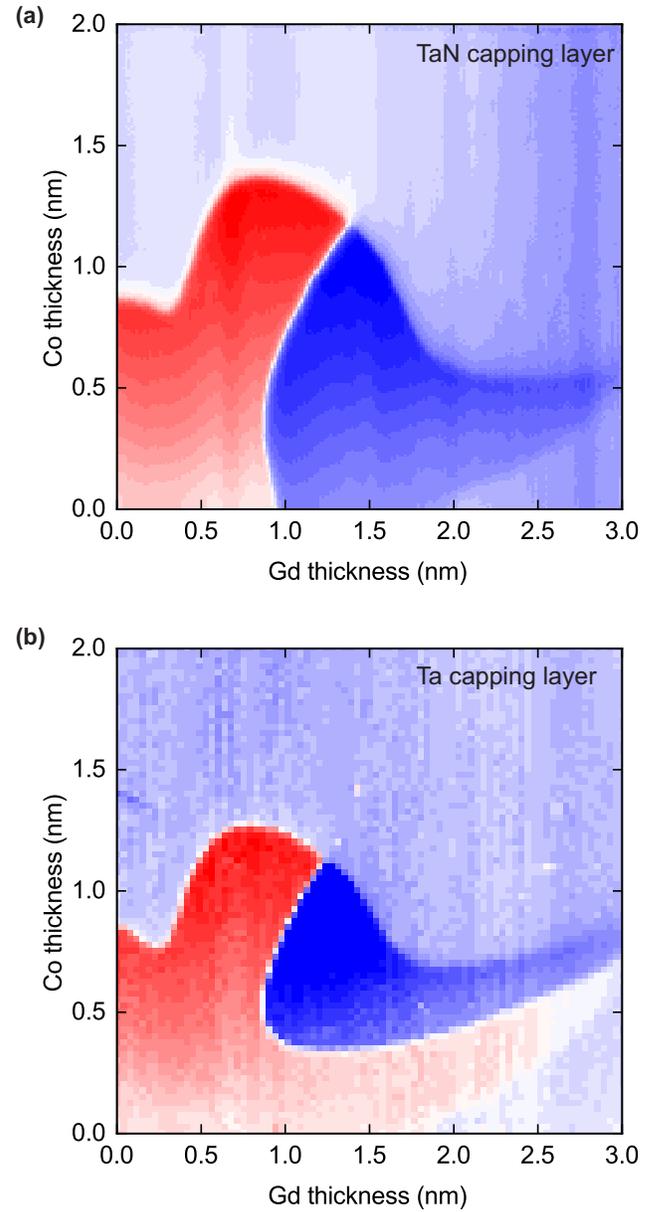}
\caption{\label{fig:cap} Magnetostatic phase diagrams of the Co(0.6)/Gd(x)/Co(0.7)/Gd(1.5) materials system with a 4 nm thick capping layer of a): TaN, and b): Ta. }
\end{figure}

In Fig.\ref{fig:cap} a and b the magnetostatic phase diagrams of a Co(0.6)/Gd(x)/Co(0.7)/Gd(1.5) stack with a 4 nm thick capping layer of TaN and Ta are plotted, respectively. The most important difference to note here is that the region where the magnetization is OOP and the Gd-contribution is dominant (dark blue) extends all the way to zero Co thickness for the TaN cap. In contrast, the Ta-capped sample magnetization only becomes dominated by the Gd magnetization for a minimum Co thickness of about 0.4 nm.
In Fig. \ref{fig:cap} a and b we plot magnetostatic phase diagrams for a Ta(4)/Pt(4)/Co(0.6)/Gd(x)/Co(0.7)/Gd(1.5) with a TaN and Ta capping layer, respectively.

\bibliographystyle{ieeetr}
\bibliography{ms.bib}

\end{document}